\documentclass[amsfonts,amssymb,amsmath,twocolumn,notitlepage,nofootinbib,%
  groupedaddress]{revtex4-1}
\usepackage[dvips]{graphicx}

\begin{document}
\title{Charmonium potentials: Matching perturbative and lattice QCD}
\author{Alexander Laschka}
\author{Norbert Kaiser}
\author{Wolfram Weise}
\affiliation{Physik Department, Technische Universit\"{a}t M\"{u}nchen,
  D-85747 Garching, Germany}
\date{August 15, 2012}

\begin{abstract}
Central and spin-spin potentials for charmonium, constructed from
Nambu-Bethe-Salpeter amplitudes in lattice simulations of full QCD, are matched
with results from perturbative QCD at an appropriate distance scale. This
matching is made possible by defining the perturbative potentials through
Fourier transforms with a low-momentum cutoff. The central (spin-independent)
potential is compared with potentials derived from an expansion in powers of
the inverse quark mass. A well-controlled continuation of the charmonium
spin-spin potential from lattice QCD to short distances is performed. The mass
splittings of the charmonium singlet and triplet states below the open charm
threshold, obtained from the matched spin-spin potential, are in good agreement
with the experimental values.
\end{abstract}

\maketitle

\section{Introduction}
Within the framework of potential non-relativistic QCD (pNRQCD) the
heavy-quark-antiquark potential is a well-defined quantity, and usually
presented as an expansion in powers of the inverse heavy-quark
mass~\cite{Brambilla:2004jw}. The leading-order static potential (i.e., the
potential between infinitely heavy quarks) is accurately known from lattice QCD
studies. The potentials at order $1/m$ and $1/m^2$ have been computed in
quenched lattice QCD using the Wilson-loop
formalism~\cite{Koma:2006si,Koma:2006fw,Koma:2007jq}. In this approach, a
spin-spin potential is found at order $1/m^2$ with a sign such that the mass
ordering of hyperfine multiplets is opposite to the ordering observed
empirically. An earlier extraction of the spin-spin potential from lattice
QCD~\cite{Bali:1997am} found a sign in agreement with the empirical ordering.

A different lattice QCD approach to extract the heavy-quark potentials has
recently been proposed in Ref.~\cite{Kawanai:2011xb}. The spin-spin and central
potentials at finite quark mass are obtained from Nambu-Bethe-Salpeter (NBS)
amplitudes through an effective Schr\"{o}dinger equation. A repulsive spin-spin
potential which shifts spin triplet states upward and spin singlet states
downward is found as required for reproducing the experimental charmonium and
bottomonium spectra. Although some model dependence is involved in this
approach, the resulting shape of the potential is quite different from a
$\delta$-function potential that one commonly obtains from one-gluon exchange
with an effective coupling strength $\alpha_s$ treated as an adjustable
parameter.

In this Letter we show that one-gluon exchange is, in fact, sufficient to
derive a perturbative spin-spin potential, consistent with the recent findings
in lattice QCD, if the running of the QCD coupling $\alpha_s(q)$ is properly
included. A matching of perturbative and non-perturbative regions of the
spin-spin potential is thus possible. The central (spin-independent) charmonium
potential, obtained in full lattice QCD from NBS amplitudes, is matched to a
corresponding perturbative potential and compared with results obtained
previously in the $1/m$ expansion~\cite{Laschka:2011zr}. The perturbative
potentials are constructed with a restricted Fourier transformation. This
method replaces the frequently used renormalon subtraction
scheme~\cite{Pineda:2001zq}. The S-wave charmonium spectrum is derived in
Section~\ref{sec:3} from the combination of the matched central and spin-spin
potentials and compared with experimental values. The single free parameter in
our approach is an overall additive constant which enters in the extraction of
the charm-quark mass.

\section{Construction and comparison of charmonium potentials}
The quark-antiquark potential is commonly written in the form
\begin{equation}
  V_{q\bar q}(r)=V_C(r)+\vec S_q\!\cdot\!\vec S_{\bar q}\ V_S(r)+\ldots\, ,
\end{equation}
with a central potential $V_C(r)$ and a spin-spin potential $V_S(r)$. First,
consider the perturbative central (spin-independent) charmonium potential. As
in Ref.~\cite{Laschka:2011zr}, we define it in $r$-space via a restricted
Fourier transformation with a low-momentum cutoff~$\mu_C$:
\begin{equation}
  \label{eq:2}
  V_C^{\text{pert}}(r,\mu_C) = \intop_{q>\mu_C}\!\!\!\frac{d^3q}{(2\pi )^3}\ 
  e^{i\vec q\cdot\vec r}\,\bigg[\tilde{V}^{(0)}(q)
  +\frac{\tilde{V}^{(1)}(q)}{m/2}\bigg]\, ,
\end{equation}
with $q=|\vec q\,|$. It includes the static potential $\tilde{V}^{(0)}(q)$ at
two-loop order in momentum space
\begin{equation}
  \label{eq:3}
  \tilde{V}^{(0)}(q) = -\frac{16\pi\alpha_s(q)}{3q^2}\,\bigg[1
  +\frac{\alpha_s(q)}{4\pi}\, a_1+\frac{\alpha_s^2(q)}{(4\pi)^2}\, a_2\bigg]
  \, ,
\end{equation}
and the $1/m$ potential at leading order~\cite{Brambilla:2000gk}:
\begin{equation}
  \tilde{V}^{(1)}(q) = -\frac{2\pi^2\alpha_s^2(q)}{q}\, .
\end{equation}
The coefficients $a_1$ and $a_2$ in Eq.~(\ref{eq:3}) are known
analytically~\cite{Peter:1996ig,Peter:1997me,Schroder:1998vy} and have the
values
\begin{align}
  a_1 &= 7\, ,\\
  a_2 &= \frac{695}{6}+36\pi^2-\frac{9}{4}\pi^4+14\zeta(3)\, ,
\end{align}
for three light quark flavors. The low-momentum region excluded in
Eq.~(\ref{eq:2}) is not accessible in perturbation theory. It is substituted by
an additive overall constant to the potential (see~\cite{Laschka:2011zr} for
details).

The full four-loop renormalization-group (RG)
running~\cite{vanRitbergen:1997va} with three light flavors is implemented for
the strong coupling $\alpha_s(q)$. We use the input value
$\alpha_s(1.25\ \text{GeV})=0.406\pm0.010$, derived from
$\alpha_s(m_Z\!=\!91.1876\ \text{GeV})=0.1184\pm 0.0007$~\cite{Bethke:2009jm},
taking into account flavor thresholds~\cite{Chetyrkin:1997sg}.

The perturbative potential is matched at a suitable distance $r_m$ to the
spin-independent potential calculated in full lattice QCD using NBS
amplitudes~\cite{Kawanai:2011jt}. This lattice potential includes charm-quark
mass effects to all orders and can be parametrized as
\begin{equation}
  V_C^{\text{lat}}(r) = -\frac{A}{r}+\sigma\, r+\text{const}\, ,
\end{equation}
with $A=0.813\pm 0.022$ and $\sqrt{\sigma}=(0.394\pm 0.007)$~GeV. Optimal
matching is achieved with a low-momentum cutoff $\mu_C=(0.54\pm 0.02)$~GeV. In
the region around the chosen matching position $r_m=0.14$~fm, the perturbative
potential and the lattice potential are both expected to be reliable. The
potential changes only marginally under limited variations of the matching
point.

\begin{figure}[tp]
  \includegraphics[width=\columnwidth]{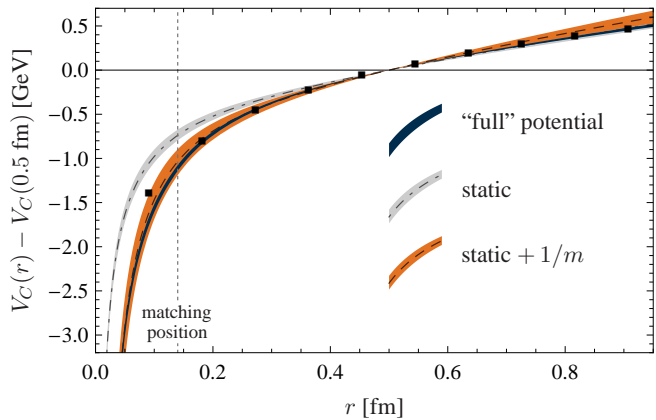}
  \caption{\label{fig:1}Charmonium potential from a combination of perturbative
  QCD and lattice QCD~\cite{Kawanai:2011jt} matched at $r_m=0.14$~fm (solid
  curve). The dot-dashed line (with error band) shows the static potential from
  Ref.~\cite{Laschka:2011zr} for comparison, while the dashed line (with error
  band) shows the static-plus-$1/m$ potential from Ref.~\cite{Laschka:2011zr}.
  The energy scale is chosen relative to the potential at $r=0.5$~fm for
  convenience.}
\end{figure}

In Fig.~\ref{fig:1} we compare this matched central potential (solid curve)
with the potential obtained in the Wilson-loop approach using a $1/m$
expansion. One observes that the ``full'' potential differs significantly from
the static potential constructed in Ref.~\cite{Laschka:2011zr} (dot-dashed line
with error band), but it is consistent within errors when the $1/m$ potential
from Ref.~\cite{Laschka:2011zr} is added to the static potential (dashed line
with error band). The charm-quark mass $m$ relevant for the $1/m$ potential has
been varied in Fig.~\ref{fig:1} in the range $(1.5\pm 0.2)$~GeV.

Let us now focus on the spin-spin potential. Tensor and spin-orbit terms are
not discussed in this Letter. These have so far not been studied in the new
lattice QCD approach employing the NBS amplitudes. The authors of
Ref.~\cite{Kawanai:2011jt} fit the spin-spin potential from full lattice QCD
with three different functional forms. We use the exponential form
\begin{equation}
  \label{eq:8}
  V_S^{\text{lat}}(r) = \alpha\, e^{-\beta r}\, ,
\end{equation}
with $\alpha=(0.825\pm 0.019)$~GeV and $\beta=(1.982\pm 0.024)$~GeV, since it
provides the best fit to the lattice data.

In the following we construct a perturbative spin-spin potential that can be
used to continue the lattice potential to short distances. Recall first the
well-known (schematic) spin-spin potential derived from one-gluon exchange
assuming a constant coupling $\alpha_s$:
\begin{equation}
  V_S(r) = \frac{32\pi}{9m_q^2}\,\alpha_s\,\delta^3(\vec r\,)\, ,
\end{equation}
with the quark/antiquark mass $m_q$. This $\delta$-function term is in
agreement with the leading-order spin-spin potential obtained in
pNRQCD~\cite{Brambilla:2004jw}.

In analogy to the case of the spin-independent potential, it is possible to
include the RG running of $\alpha_s(q)$ in the construction of the spin-spin
potential. We define the perturbative part of $V_S(r)$ as:
\begin{equation}
  V_S^{\text{pert}}(r,\mu_S) = \frac{32\pi}{9m^2}\intop_{q>\mu_S}\!\!\!
  \frac{d^3q}{(2\pi )^3}\ e^{i\vec q\cdot\vec r}\alpha_s(q)\, ,
\end{equation}
with a low-momentum cutoff $\mu_S$. For $q>\mu_S$ the perturbative RG evolution
of $\alpha_s(q)$ is supposed to be reliable. The non-perturbative infrared
behavior of the quark and gluon couplings prohibits a controlled low-momentum
extension for $q<\mu_S$. It is nevertheless useful to examine such an extension
for $r\ll 1/\mu_S$ in the form
\begin{equation}
  \label{eq:11}
  V_S^{\text{ir}} = \frac{32\pi}{9m^2}\,\overline \alpha_s\!\!\!
  \intop_{q<\mu_S}\!\!\!\frac{d^3q}{(2\pi )^3}\, e^{i\vec q\cdot\vec r}\simeq
  \frac{16\,\overline\alpha_s}{27\pi m^2}\,\mu_S^3\, ,
\end{equation}
with a parameter $\overline \alpha_s$ reflecting the average interaction
strength in the infrared region. Note that Eq.~(\ref{eq:11}) gives a positive
constant proportional to $\mu_S^3$ to be added as a correction to
$V_S^{\text{pert}}(r,\mu_S)$, with $\overline \alpha_s$ not known, but expected
to be of $\mathcal O(1)$.

In order to facilitate the numerical evaluation of the Fourier integral for the
perturbative part, it is useful to rewrite $V_S^{\text{pert}}(r,\mu_S)$ in the
form
\begin{equation}
  \label{eq:12}
  V_S^{\text{pert}}(r,\mu_S) = -\frac{16}{9m^2\pi r}\,
  \frac{\partial}{\partial r}\intop_{\mu_S}^{\infty}\!dq\cos(qr)\,\alpha_s(q)
  \, .
\end{equation}

\begin{figure}[tp]
  \includegraphics[width=\columnwidth]{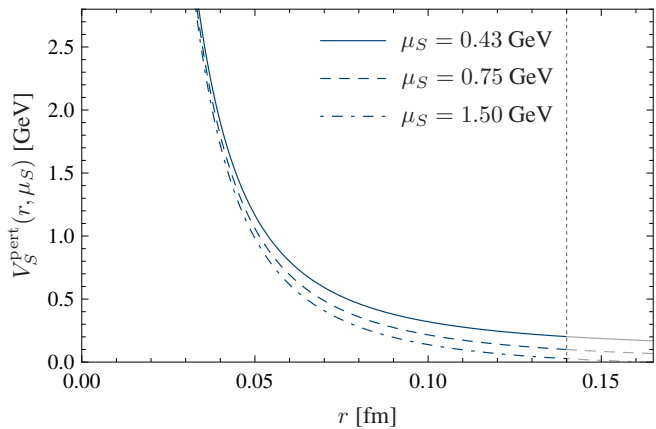}
  \caption{\label{fig:2}Spin-spin potential in $r$-space derived from
  one-gluon exchange as defined in Eq.~(\ref{eq:12}). The potential is shown
  for different values of the low-momentum cutoff $\mu_S$.}
\end{figure}

To be consistent with the lattice QCD analysis, the charm-quark mass $m$ in the
denominator of Eq.~(\ref{eq:12}) is identified with $(1.74\pm 0.03)$~GeV, the
kinetic quark mass that has been determined in Ref.~\cite{Kawanai:2011jt}.
Discarding the additive constant $V_S^{\text{ir}}$ in the first step, the
resulting short distance potential and its dependence on the cutoff scale
$\mu_S$ is shown in Fig.~\ref{fig:2}. The shape is evidently very different
from a Gaussian or a $\delta$-function. Such forms are frequently used for the
spin-spin potential in phenomenological models.

\begin{figure}[tp]
  \includegraphics[width=\columnwidth]{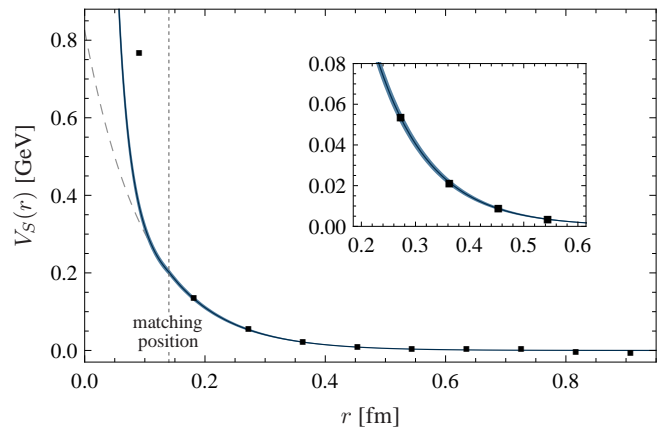}
  \caption{\label{fig:3}Charmonium spin-spin potential (with error band) from a
  combination of perturbative QCD (see Eq.~(\ref{eq:12})) and lattice
  QCD~\cite{Kawanai:2011jt}, matched at $r_m=0.14$~fm. Dashed line:
  continuation of the exponential lattice fit (\ref{eq:8}) to short distances.}
\end{figure}

The two components of the spin-spin potential, arising from perturbative QCD
for $r\leq r_m$ and from lattice QCD for $r\geq r_m$, can be matched at
$r_m=0.14$~fm. When ignoring the additive constant $V_S^{\text{ir}}$ this
matching is achieved for $\mu_S=(0.43\pm 0.02)$~GeV. With inclusion of the
(variable) additive constant $V_S^{\text{ir}}$ the infrared cutoff $\mu_S$ can
be varied over a wide range almost without any effect on the spin-spin
potential. Figure~\ref{fig:3} shows the matched potential with an infrared
cutoff of $\mu_S=(0.75\pm 0.25)$~GeV.

\section{Spectroscopy and charm-quark mass}
\label{sec:3}
Here, we discuss the charmonium spectrum below the
$\text{D}\overline{\text{D}}$ threshold as derived from the matched potentials
constructed in the previous section. We focus on the 1S and 2S states which are
not influenced by the tensor and spin-orbit interactions. The Schr\"{o}dinger
equation for these states,
\begin{multline}
  \label{eq:13}
  \bigg[-\frac{\vec\nabla^2}{m}+2m_{\text{PS}}(\mu_C)+V_C(r)\\
  +\vec S_q\!\cdot\!\vec S_{\bar q}\ V_{S}(r)-E\bigg]\psi(\vec r\,) = 0\, ,
\end{multline}
involves a single free parameter $m_{\text{PS}}(\mu_C)$, the
($\mu_C$-dependent) charm-quark mass in the potential subtracted (PS)
scheme~\cite{Beneke:1998rk}. In order to make the PS scheme applicable for our
purposes, we extend it by including the $1/m$-correction term:
\begin{multline}
  m_{\text{PS}}(\mu_C) = m_{\text{pole}}\\
  +\frac{1}{2}\intop_{q<\mu_C}\!\!\!\frac{d^3q}{(2\pi )^3}\,\bigg[
  \tilde{V}^{(0)}(q)+\frac{\tilde{V}^{(1)}(q)}{m/2}\bigg].
\end{multline}

Replacing the operator $\vec S_q\!\cdot\!\vec S_{\bar q}$ by its eigenvalues,
$-3/4$ for the spin singlet and $1/4$ for the spin triplet, the Schr\"{o}dinger
equation~(\ref{eq:13}) is solved numerically. The spin-spin potential $V_S(r)$
can be included in two different ways. In the first case it is treated in
first-order perturbation theory, in the second case it is fully included in the
Schr\"{o}dinger equation. Due to the singular behavior of $V_S(r)\sim r^{-2.8}$
for $r\to 0$ (according to our construction) the wave functions diverge
(mildly) in the latter case for very small values of $r$. In this case we solve
the radial Schr\"{o}dinger equation numerically for $r>0.003$~fm and convince
ourselves that the physical results are not affected by contributions coming
from shorter distances. The results of the two alternative treatments of the
spin-spin potential agree within $12$~MeV (see Table~\ref{table:1}). The
calculated mass splittings between singlet and triplet states are in good
agreement with experimental results for both 1S and 2S charmonia.

\begin{table}[bp]
  \begin{tabular*}{\columnwidth}{@{\extracolsep{\fill}}lccc}
  \hline
  \hline
  &Case 1&Case 2&Experiment~\cite{Nakamura:2010zzi}\\
  \hline
  1S mass splitting [MeV]&$117\pm 6$&$105\pm 6$&$116.6\pm 1.2$\\
  2S mass splitting [MeV]&$56\pm 3$&$46\pm 3$&$49\pm 4$\\
  \hline
  \hline
  \end{tabular*}
  \caption{\label{table:1}Predicted mass splittings of charmonium 1S and 2S
  multiplets in comparison with experimental data.
  Case 1: spin-spin potential treated in first-order perturbation theory.
  Case 2: spin-spin potential fully included in the Schr\"{o}dinger equation.}
\end{table}

\begin{figure}[tp]
  \includegraphics[width=0.912\columnwidth]{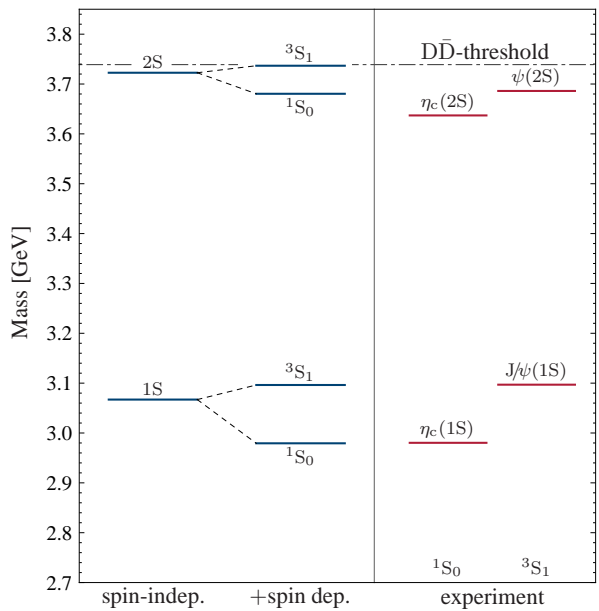}
  \caption{\label{fig:4}Predicted masses of charmonium 1S and 2S states in
  comparison with experimental data~\cite{Nakamura:2010zzi}. The results in the
  first column are based on the spin-independent potential only. The effects of
  the spin-spin potential, treated in first-order perturbation theory, are
  added in the second column.
}
\end{figure}

The single free parameter $m_{\text{PS}}(\mu_C)$ is chosen such that the
spin-weighted average of the 1S states agrees with its experimental
value~\cite{Nakamura:2010zzi}. For the excited 2S states our approach predicts
masses for $\eta_c(\text{2S})$ and $\psi(\text{2S})$ that are slightly too
large (see Fig.~\ref{fig:4}). However, these states are close to the
$\text{D}\overline{\text{D}}$ threshold. Going beyond this threshold requires a
complex (energy-dependent) $\text{c}\overline{\text{c}}$ potential or an
explicit treatment of coupled channels. While the imaginary part starts at the
opening of the $\text{D}\overline{\text{D}}$ channel, the corresponding
dispersive real part induces an attractive shift of the 2S states. In
second-order perturbation theory this shift is proportional to the squared
$\text{c}\overline{\text{c}}\to\text{D}\overline{\text{D}}$ transition matrix
element.

The predicted size of the hyperfine splittings is quite sensitive to the value
of the mass $m$ in the denominator of Eq.~(\ref{eq:12}). For example, choosing
$m=1.5$~GeV instead of the kinetic quark mass
($1.74\pm 0.03)$~GeV~\cite{Kawanai:2011jt} would give rise to a 1S mass
splitting that is about $20\%$ too large. This is in contrast to variations of
the matching position $r_m$ and the infrared cutoff $\mu_S$, which affect the
hyperfine splittings only marginally.

The spin-spin potential, as constructed in the previous section, produces a
non-vanishing but small splitting between the 1P singlet and triplet states,
namely $\text{h}_{\text c}(\text{1P})$ and $\chi_{\text{cj}}(\text{1P})$,
unlike the $\delta$-function spin-spin potential. In first-order perturbation
theory the effect amounts to a mass difference of $(8.2\pm 0.5)$~MeV. The full
inclusion of the spin-spin potential $V_S(r)$ in the Schr\"{o}dinger equation
gives rise to a slightly larger mass splitting of $(8.3\pm 0.5)$~MeV.

The value of the mass parameter $m_{\text{PS}}(\mu_C)$, determined in our
approach by fitting to empirical charmonium spectra, can be translated into
alternative schemes for quark masses. The PS mass $m_{\text{PS}}(\mu_C)$ is
first converted to the pole mass and in a second step mapped onto the
$\overline{\text{MS}}$ mass
$\overline m_c\equiv m_{\overline{\text{MS}}}(m_{\overline{\text{MS}}})$. This
procedure is described in detail in Ref.~\cite{Laschka:2011zr}. Applying the
same method here, we find for the charm-quark mass in the
$\overline{\text{MS}}$ scheme
\begin{equation}
  \overline m_c=(1.21\pm 0.04)\text{ GeV},
\end{equation}
in good agreement with other
determinations~\cite{Laschka:2011zr,Nakamura:2010zzi}. The error reflects
combined uncertainties in the lattice potentials, in the input value of the
strong coupling $\alpha_s(q)$, and from the matching to the empirical states.

We close with a few remarks concerning bottomonium: until now, the bottomonium
spin-spin potential has not been studied within the new lattice QCD approach
based on NBS amplitudes. An extrapolation of the spin-spin potential from
charmonium to bottomonium can be done by simply assuming a $1/m^2$ dependence
of the lattice potential and allowing for variations of the mass parameter $m$.
In the perturbative part of the potential we account furthermore for a modified
running of $\alpha_s(q)$ due to four massless flavors and use
$\alpha_s(4.2\text{ GeV})=0.226\pm 0.003$ as an input. The empirical mass
splitting of $(69\pm 3)$~MeV~\cite{Nakamura:2010zzi} between
$\eta_{\text b}(\text{1S})$ and $\Upsilon(\text{1S})$ can be reproduced either
for a kinetic bottom-quark mass $m=4.7$~GeV (with the spin-spin potential
treated in first-order perturbation theory), or with $m=4.3$~GeV (if the
spin-spin potential is fully included in the Schr\"{o}dinger equation). It will
be interesting to have available the corresponding lattice QCD results for
bottomonium.

\section{Summary}
Central and spin-spin potentials for charmonium have been derived by combining
perturbative QCD at small distances ($r<0.14$~fm) with results from lattice QCD
for larger distances up to $r\simeq 1$~fm. By defining the perturbative
potentials via a restricted Fourier transformation this matching has been made
possible. We have found that the central quark-antiquark potential, constructed
from NBS amplitudes in full QCD lattice
simulations~\cite{Kawanai:2011xb,Kawanai:2011jt}, agrees within errors with the
static-plus-$1/m$ potential derived in the Wilson-loop
formalism~\cite{Koma:2006si,Koma:2006fw,Koma:2007jq}. The matched spin-spin
potential produces hyperfine splittings for the S-wave charmonium states that
are in good agreement with experiment. The $\overline{\text{MS}}$ mass of the
charm quark also agrees well with other determinations of this QCD parameter.

\begin{acknowledgments}
This work was supported in part by BMBF, GSI and the DFG Excellence Cluster
``Origin and Structure of the Universe''. We thank Antonio Vairo for useful
discussions. One of the authors (A.\,L.) acknowledges partial support by the
TUM Graduate School.
\end{acknowledgments}

\end{document}